\documentstyle[amssymb,preprint,prb,aps]{revtex}

\begin{document}

\begin{titlepage}

\title
{Spin tunneling properties in mesoscopic magnets:
effects of a magnetic field}

\author{Rong L\"{u},\footnote {Author to whom 
the correspondence should be addressed.\\
Electronic address: rlu@castu.tsinghua.edu.cn} Hui Pan,
Jia-Lin Zhu, and Bing-Lin Gu} 
\address{Center for Advanced Study, Tsinghua University,
Beijing 100084, P. R. China}

\date{\today}

\maketitle
\begin{abstract}
The tunneling of a giant spin at excited levels
is studied theoretically in
mesoscopic magnets with
a magnetic field at an arbitrary angle in the easy plane.
Different structures of the tunneling barriers
can be generated by the magnetocrystalline
anisotropy, the magnitude and the orientation
of the field.
By calculating the nonvacuum instanton solution explicitly, 
we obtain the tunnel splittings and the
tunneling rates for different angle ranges of the external magnetic field
($\theta_{H}=\pi/2$ and $\pi/2<\theta_{H}<\pi$).
The temperature dependences of
the decay rates are clearly shown for each case.
It is found that the tunneling rate and
the crossover temperature depend on the 
orientation of the external magnetic
field. This feature can be tested with the
use of existing experimental techniques.

\noindent
{\bf PACS number(s)}:  75.45.+j, 75.50.Jm
\end{abstract}

\end{titlepage}

\section*{I. Introduction}

Recently, nanospin systems have emerged as good candidates to display
quantum phenomena at a mesoscopic or macroscopic scale.\cite{1} Theoretical
investigations showed that quantum tunneling was possible in ferromagnetic
(FM) nanoparticles containing as much as $10^5-10^6$ spins.\cite{1} At
extremely low temperature, the magnitude of the total magnetization ${\bf M}$
is frozen out and thereby its direction becomes the only dynamical variable.
In the absence of an external magnetic field, the magnetocrystalline
anisotropy can create energetically degenerate easy directions depending on
the crystal symmetry. Tunneling between neighboring states removes the
degeneracy of the original ground states and leads to a level splitting.
This phenomenon is called macroscopic quantum coherence (MQC). However, MQC
is hard to be observed in experiments without controlling the height and the
width of the barrier. It has been believed that a magnetic field is a good
external parameter to make the quantum tunneling observable. By applying a
magnetic field in a proper direction, one of the two energetically
equivalent orientations becomes metastable and the magnetization vector can
escape from the metastable state through the barrier to a stable one, which
is called macroscopic quantum tunneling (MQT). A large number of experiments
involving resonance measurements, magnetic relaxation, and hysteresis loop
study, M\"{o}ssbauer spectroscopy, and neutron scattering study for various
systems showed either temperature-independent relaxation phenomena or a
well-defined resonance depending exponentially on the number of total spins,
which supported the idea of magnetic quantum tunneling.\cite{1}

To our knowledge, the tunneling of a single spin degree of freedom was first
studied by Korenblit and Shender in 1978.\cite{1.5} More recently, the
tunneling problem of the magnetization reversal was studied extensively for
the single-domain FM nanoparticles in a magnetic field applied at an
arbitrary angle. This problem was studied by Zaslavskii with the help of
mapping the spin system onto a one-dimensional particle system.\cite{2} For
the same system, Miguel and Chudnovsky\cite{3} calculated the tunneling rate
by applying the imaginary-time path integral, and demonstrated that the
angular and field dependences of the tunneling exponent obtained by
Zaslavskii's method and by the path-integral method coincide precisely. Kim
and Hwang performed a calculation based on the instanton technique for FM
particles with biaxial and tetragonal crystal symmetry,\cite{4} and Kim
extended the tunneling rate for biaxial crystal symmetry to a finite
temperature.\cite{5} The quantum-classical transition of the escape rate for
uniaxial spin system in an arbitrarily directed field was investigated by
Garanin, Hidalgo and Chudnovsky with the help of mapping onto a particle
moving in a double-well potential.\cite{6} The switching field measurement
was carried out on single-domain FM nanoparticles of Barium ferrite
(BaFeCoTiO) containing about $10^5-10^6$ spins.\cite{7} The measured angular
dependance of the crossover temperature was found to be in excellent
agreement with the theoretical prediction,\cite{3} which strongly suggests
the quantum tunneling of magnetization in the BaFeCoTiO nanoparticles.
L\"{u} {\it et al}. studied the quantum tunneling of the N\'{e}el vector in
single-domain antiferromagnetic (AFM) nanoparticles with biaxial,
tetragonal, and hexagonal crystal symmetry in an arbitrarily directed field.%
\cite{8}

It is noted that the previous results of spin tunneling at excited levels in
an arbitrarily directed field were obtained by numerically solving the
equation of motion satisfied by the least trajectory,\cite{5} and the system
considered in Ref. 6 had the simple biaxial crystal symmetry. The purpose of
this paper is to present an analytical study of the quantum tunneling {\it %
at excited levels} in the FM particles with an arbitrarily directed field.
Moreover, the system considered in this paper has a much more complex
structure (i. e., {\it the general structure in experiments}), such as
trigonal, tetragonal, and hexagonal crystal symmetry. By applying an
arbitrarily directed magnetic field, the problem does not possess any
symmetry and for that reason is more difficult mathematically. However, it
is worth pursuing because of its significance for experiments and the
easiest to implement in practice. Since the result of spin tunneling at
excited levels for tetragonal symmetry is a generalization of that of
tunneling at ground-state levels studied by Kim and Hwang,\cite{4} we can
compare our results with theirs by taking the low-energy limit. We will show
that MQC and MQT can be consecutively observed by changing the direction of
magnetic field, and discuss their dependence on the direction and the
magnitude of field. The dependence of the crossover temperature $T_c$ and
the magnetic viscosity (which is the inverse of WKB\ exponent at the
quantum-tunneling-dominated regime $T\ll T_c$) on the direction and the
magnitude of the field, and the magnetic anisotropies is expected to be
observed in future experiments on individual single-domain particles with
different magnetocrystalline anisotropies. Both the nonvacuum (or thermal)
instanton or bounce solution, the WKB exponents and the preexponential
factors are evaluated exactly for different angle ranges of the magnetic
field $(\theta _H=\pi /2$ and $\pi /2<\theta _H<\pi )$. The low-energy limit
of our results agrees well with that of ground-state spin tunneling. In
order to compare theories with experiments, predictions of the crossover
temperature corresponding to the transition from classical to quantum
behavior and the temperature dependence of the decay rate are clearly shown
in this paper. Both variables are expressed as a function of parameters
which can be changed experimentally, such as the number of total spins, the
effective anisotropy constants, the strength and orientation of applied
magnetic field. Our results show that the distinct angular dependence,
together with the dependence of the WKB tunneling rate on the strength of
the external magnetic field, may provide an independent experimental test
for the spin tunneling at excited levels in nanoscale magnets. When the
effective magnetic anisotropy of the particle is known, our theoretical
results give clear predictions with no fitting parameters. Therefore,
quantum spin tunneling could be studied as a function of the effective
magnetic anisotropy. Our results should be helpful for future experiments on
spin tunneling in single-domain particles with different magnetocrystalline
anisotropies.

This paper is structured in the following way. In Sec. II, we review briefly
some basic ideas of spin tunneling in FM\ particles. And we discuss the
fundamentals concerning the computation of level splittings and tunneling
rates of excited states in the double-well-like potential. In Secs. III, IV,
and V, we study the spin tunneling at excited levels for FM particles with
trigonal, tetragonal and hexagonal crystal symmetry in an external magnetic
field applied in the $ZX$ plane with a range of angles $\pi /2\leq \theta
_H<\pi $, respectively. The conclusions are presented in Sec. VI.

\section*{II. Physical model of spin tunneling in FM particles}

For a spin tunneling problem, the tunnel splitting or the tunneling rate is
determined by the imaginary-time transition amplitude from an initial state $%
\left| i\right\rangle $ to a final state $\left| f\right\rangle $ as 
\begin{equation}
{\cal U}_{fi}=\left\langle f\right| e^{-{\cal H}T}\left| i\right\rangle
=\int {\cal D}\Omega \exp \left( -{\cal S}_E\right) ,  \eqnum{1}
\end{equation}
where ${\cal S}_E$ is the Euclidean action and ${\cal D}\Omega $ is the
measure of the path integral. In the spin-coherent-state representation, the
Euclidean action is 
\begin{equation}
{\cal S}_E\left( \theta ,\phi \right) =\frac V\hbar \int d\tau \left[ i\frac{%
M_0}\gamma \left( \frac{d\phi }{d\tau }\right) -i\frac{M_0}\gamma \left( 
\frac{d\phi }{d\tau }\right) \cos \theta +E\left( \theta ,\phi \right)
\right] ,  \eqnum{2}
\end{equation}
where $V$ is the volume of the FM particle and $\gamma $ is the gyromagnetic
ratio. $M_0=\left| {\bf M}\right| =\hbar \gamma S/V$, where $S$ is the total
spin of FM particles. It is noted that the first two terms in Eq. (2) define
the Berry phase or Wess-Zumino, Chern-Simons term which arises from the
nonorthogonality of spin coherent states in the north-pole parametrization.
The Wess-Zumino term has a simple topological interpretation. For a closed
path, this term equals $-iS$ times the area swept out on the unit sphere
between the path and the north pole. The first term in Eq. (2) is a total
imaginary-time derivative, which has no effect on the classical equations of
motion, but it is crucial for the spin-parity effects.\cite{1,9,10,11,12,13}
However, for the closed instanton or bounce trajectory described in this
paper (as shown in the following), this time derivative gives a zero
contribution to the path integral, and therefore can be omitted.

In the semiclassical limit, the dominant contribution to the transition
amplitude comes from finite action solution (instanton or bounce) of the
classical equation of motion. The instanton's contribution to the tunneling
rate $\Gamma $ or the tunnel splitting $\Delta $ is given by\cite{1} 
\begin{equation}
\Gamma \ (\text{or }\ \Delta )=A\omega _p\left( \frac{{\cal S}_{cl}}{2\pi }%
\right) ^{1/2}e^{-{\cal S}_{cl}},  \eqnum{3}
\end{equation}
where $\omega _p$ is the oscillation frequency in the well, ${\cal S}_{cl}$
is the classical action, and the prefactor $A$ originates from the quantum
fluctuations about the classical path. It is noted that Eq. (3) is based on
quantum tunneling at the level of ground state, and the temperature
dependence of the tunneling rate (i.e., tunneling at excited levels) is not
taken into account. However, the instanton technique is suitable only for
the evaluation of the tunneling rate or the tunnel splitting at the vacuum
level, since the usual (vacuum) instantons satisfy the vacuum boundary
conditions. In this paper, we will calculate the nonvacuum instantons
corresponding to quantum tunneling at excited levels.

For a particle moving in a double-well-like potential $U\left( x\right) $,
the level splittings of degenerate excited levels or the imaginary parts of
the metastable levels at an energy $E>0$ are given by the following formula
in the WKB approximation,\cite{14} 
\begin{equation}
\Delta E\text{ }\left( \text{or }%
\mathop{\rm Im}%
E\right) =\frac{\omega \left( E\right) }\pi \exp \left[ -{\cal S}\left(
E\right) \right] ,  \eqnum{4}
\end{equation}
and the imaginary-time action is 
\begin{equation}
{\cal S}\left( E\right) =2\sqrt{2m}\int_{x_1\left( E\right) }^{x_2\left(
E\right) }dx\sqrt{U\left( x\right) -E},  \eqnum{5}
\end{equation}
where $x_{1,2}\left( E\right) $ are the turning points for the particle
oscillating inside the inverted potential $-U\left( x\right) $. $\omega
\left( E\right) =2\pi /t\left( E\right) $ is the energy-dependent frequency,
and $t\left( E\right) $ is the period of the real-time oscillation in the
potential well, 
\begin{equation}
t\left( E\right) =\sqrt{2m}\int_{x_3\left( E\right) }^{x_4\left( E\right) }%
\frac{dx}{\sqrt{E-U\left( x\right) }},  \eqnum{6}
\end{equation}
where $x_{3,4}\left( E\right) $ are the turning points for the particle
oscillating inside the potential $U\left( x\right) $.

\section*{III. MQC and MQT for trigonal crystal symmetry}

In this section, we study the quantum tunneling of the magnetization vector
in single-domain FM nanoparticles with trigonal crystal symmetry. The
external magnetic field is applied in the $ZX$ plane, at an angle in the
range of $\pi /2\leq \theta _H<\pi $. Now the total energy $E\left( \theta
,\phi \right) $ can be written as 
\begin{equation}
E\left( \theta ,\phi \right) =K_1\sin ^2\theta -K_2\sin ^3\theta \cos \left(
3\phi \right) -M_0H_x\sin \theta \cos \phi -M_0H_z\cos \theta +E_0, 
\eqnum{7}
\end{equation}
where $K_1$ and $K_2$ are the magnetic anisotropy constants satisfying $%
K_1\gg K_2>0$, and $E_0$ is a constant which makes $E\left( \theta ,\phi
\right) $ zero at the initial orientation. As the magnetic field is applied
in the $ZX$ plane, $H_x=H\sin \theta _H$ and $H_z=H\cos \theta _H$, where $H$
is the magnitude of the field and $\theta _H$ is the angle between the
magnetic field and the $\widehat{z}$ axis.

By introducing the dimensionless parameters as 
\begin{equation}
\overline{K}_2=K_2/2K_1,\overline{H}_x=H_x/H_0,\overline{H}_z=H_z/H_0, 
\eqnum{8}
\end{equation}
Eq. (7) can be rewritten as 
\begin{equation}
\overline{E}\left( \theta ,\phi \right) =\frac 12\sin ^2\theta -\overline{K}%
_2\sin ^3\theta \cos \left( 3\phi \right) -\overline{H}_x\sin \theta \cos
\phi -\overline{H}_z\cos \theta +\overline{E}_0,  \eqnum{9}
\end{equation}
where $E\left( \theta ,\phi \right) =2K_1\overline{E}\left( \theta ,\phi
\right) $, and $H_0=2K_1/M_0$. At finite magnetic field, the plane given by $%
\phi =0$ is the easy plane, on which $\overline{E}\left( \theta ,\phi
\right) $ reduces to 
\begin{equation}
\overline{E}\left( \theta ,\phi =0\right) =\frac 12\sin ^2\theta -\overline{K%
}_2\sin ^3\theta -\overline{H}\cos \left( \theta -\theta _H\right) +%
\overline{E}_0.  \eqnum{10}
\end{equation}
We denote $\theta _0$ to be the initial angle and $\theta _c$ the critical
angle at which the energy barrier vanishes when the external magnetic field
is close to the critical value $\overline{H}_c\left( \theta _H\right) $ (to
be calculated in the following). Then, the initial angle $\theta _0$
satisfies $\left[ d\overline{E}\left( \theta ,\phi =0\right) /d\theta
\right] _{\theta =\theta _0}=0$, the critical angle $\theta _c$ and the
dimensionless critical field $\overline{H}_c$ satisfy both $\left[ d%
\overline{E}\left( \theta ,\phi =0\right) /d\theta \right] _{\theta =\theta
_c,\overline{H}=\overline{H}_c}=0$ and $\left[ d^2\overline{E}\left( \theta
,\phi =0\right) /d\theta ^2\right] _{\theta =\theta _c,\overline{H}=%
\overline{H}_c}=0$. After some algebra, $\overline{H}_c\left( \theta
_H\right) $ and $\theta _c$ are found to be 
\begin{eqnarray}
\overline{H}_c &=&\frac 1{\left[ \left( \sin \theta _H\right) ^{2/3}+\left|
\cos \theta _H\right| ^{2/3}\right] ^{3/2}}\left[ 1-3\overline{K}_2\frac 1{%
\left( 1+\left| \cot \theta _H\right| ^{2/3}\right) ^{1/2}}\right.  \nonumber
\\
&&\left. +6\overline{K}_2\frac 1{\left( 1+\left| \cot \theta _H\right|
^{2/3}\right) ^{3/2}}\right] ,  \eqnum{11a} \\
\sin ^2\theta _c &=&\frac 1{1+\left| \cot \theta _H\right| ^{2/3}}\left[ 1-2%
\overline{K}_2\frac{\left| \cot \theta _H\right| ^{2/3}}{\left( 1+\left|
\cot \theta _H\right| ^{2/3}\right) ^{1/2}}-4\overline{K}_2\frac{\left| \cot
\theta _H\right| ^{2/3}}{\left( 1+\left| \cot \theta _H\right| ^{2/3}\right)
^{3/2}}\right] .  \eqnum{11b}
\end{eqnarray}

Now we consider the limiting case that the external magnetic field is
slightly lower than the critical field, i.e., $\epsilon =1-\overline{H}/%
\overline{H}_c\ll 1$. At this practically interesting situation, the barrier
height is low and the width is narrow, and therefore the tunneling rate in
MQT or the tunnel splitting in MQC is large. Introducing $\eta \equiv \theta
_c-\theta _0$ $\left( \left| \eta \right| \ll 1\text{ in the limit of }%
\epsilon \ll 1\right) $, expanding $\left[ d\overline{E}\left( \theta ,\phi
=0\right) /d\theta \right] _{\theta =\theta _0}=0$ about $\theta _c$, and
using the relations $\left[ d\overline{E}\left( \theta ,\phi =0\right)
/d\theta \right] _{\theta =\theta _c,\overline{H}=\overline{H}_c}=0$ and $%
\left[ d^2\overline{E}\left( \theta ,\phi =0\right) /d\theta ^2\right]
_{\theta =\theta _c,\overline{H}=\overline{H}_c}=0$, we obtain the
approximation equation for $\eta $ in the order of $\epsilon ^{3/2}$, 
\begin{eqnarray}
&&\left. -\epsilon \overline{H}_c\sin \left( \theta _c-\theta _H\right)
-\eta ^2\left( \frac 34\sin 2\theta _c+3\overline{K}_2\cos 3\theta _c\right)
\right.  \nonumber \\
&&\left. +\eta \left[ \epsilon \overline{H}_c\cos \left( \theta _c-\theta
_H\right) +\eta ^2\left( \frac 12\cos 2\theta _c-3\overline{K}_2\sin 3\theta
_c\right) \right] =0.\right.  \eqnum{12}
\end{eqnarray}
Then $\overline{E}\left( \theta ,\phi \right) $ reduces to the following
equation in the limit of small $\epsilon $, 
\begin{equation}
\overline{E}\left( \delta ,\phi \right) =2\overline{K}_2\sin ^2\left( 3\phi
/2\right) \sin ^3\left( \theta _0+\delta \right) +\overline{H}_x\sin \left(
\theta _0+\delta \right) \left( 1-\cos \phi \right) +\overline{E}_1\left(
\delta \right) ,  \eqnum{13}
\end{equation}
where $\delta \equiv \theta -\theta _0$ $\left( \left| \delta \right| \ll 1%
\text{ in the limit of }\epsilon \ll 1\right) $, and $\overline{E}_1\left(
\delta \right) $ is a function of only $\delta $ given by 
\begin{eqnarray}
\overline{E}_1\left( \delta \right) &=&-\frac 12\left[ \overline{H}_c\sin
\left( \theta _c-\theta _H\right) -\overline{K}_2\left( \cos ^3\theta _c-%
\frac 32\sin ^2\theta _c\cos \theta _c\right) \right] \left( 3\delta ^2\eta
-\delta ^3\right)  \nonumber \\
&&-\frac 12\left[ \overline{H}_c\cos \left( \theta _c-\theta _H\right) -3%
\overline{K}_2\left( \sin ^3\theta _c-4\sin \theta _c\cos ^2\theta _c\right)
\right] \left[ \delta ^2\left( \epsilon -\frac 32\eta ^2\right) +\delta
^3\eta -\frac 14\delta ^4\right]  \nonumber \\
&&-\frac 32\overline{K}_2\left( \sin ^3\theta _c-4\sin \theta _c\cos
^2\theta _c\right) \delta ^2\epsilon .  \eqnum{14}
\end{eqnarray}

In the following, we will investigate the tunneling behaviors of the
magnetization vector at excited levels in FM particles with trigonal crystal
symmetry at different angle ranges of the external magnetic field as $\theta
_H=\pi /2$ and $\pi /2<\theta _H<\pi $, respectively.

\subsection*{A. $\theta _H=\pi /2$}

For $\theta _H=\pi /2$, we have $\theta _c=\pi /2$ from Eq. (11b) and $\eta =%
\sqrt{2\epsilon }\left( 1+\frac 92\overline{K}_2\right) $ from Eq. (12). Eqs
(13) and (14) show that $\phi $ is very small for the full range of angles $%
\pi /2\leq \theta _H<\pi $. Performing the Gaussian integration over $\phi $%
, we can map the spin system onto a particle moving problem in the
one-dimensional potential well. Now the imaginary-time transition amplitude
Eqs. (1) and (2) becomes 
\begin{eqnarray}
{\cal U}_{fi} &=&\int d\delta \exp \left( -{\cal S}_E\left[ \delta \right]
\right) ,  \nonumber \\
&=&\int d\delta \exp \left\{ -\int d\tau \left[ \frac 12m\left( \frac{%
d\delta }{d\tau }\right) ^2+U\left( \delta \right) \right] \right\} , 
\eqnum{15}
\end{eqnarray}
with the effective mass 
\[
m=\frac{\hbar S^2}{2VK_1\left[ 1-\epsilon +9\overline{K}_2\right] }, 
\]
and the effective potential 
\begin{equation}
U\left( \delta \right) =\frac{K_1V}{4\hbar }\delta ^2\left( \delta -2\eta
\right) ^2.  \eqnum{16}
\end{equation}
The plot of the effective potential $\overline{E}_1\left( \delta \right) $
as a function of $\delta \left( =\theta -\theta _0\right) $ for $\theta
_H=\pi /2$ is shown in Fig. 1, and $\hbar U\left( \delta \right) =2K_1V%
\overline{E}_1\left( \delta \right) $. The problem is one of MQC, where the
magnetization vector resonates coherently between the energetically
degenerate easy directions at $\delta =0$ and $\delta =2\sqrt{2\epsilon }%
\left( 1+\frac 92\overline{K}_2\right) $ separated by a classically
impenetrable barrier at $\delta =\sqrt{2\epsilon }\left( 1+\frac 92\overline{%
K}_2\right) $.

The nonvacuum (or thermal) instanton configuration $\delta _p$ which
minimizes the Euclidean action in Eq. (16) satisfies the equation of motion 
\begin{equation}
\frac 12m\left( \frac{d\delta _p}{d\tau }\right) ^2-U\left( \delta _p\right)
=-E,  \eqnum{17}
\end{equation}
where $E>0$ is a constant of integration, which can be viewed as the
classical energy of the pseudoparticle configuration. Then the kink-solution
is 
\begin{equation}
\delta _p=\eta +\sqrt{\eta ^2-\alpha }\text{sn}\left( \omega _1\tau
,k\right) ,  \eqnum{18}
\end{equation}
where $\alpha =2\sqrt{\frac{\hbar E}{K_1V}}$, and $\omega _1=\sqrt{\frac{K_1V%
}{2\hbar m}}\sqrt{\eta ^2+\alpha }$. sn$\left( \omega _1\tau ,k\right) $ is
the Jacobian elliptic sine function of modulus $k=\sqrt{\frac{\eta ^2-\alpha 
}{\eta ^2+\alpha }}$. The Euclidean action of the nonvacuum instanton
configuration Eq. (18) over the domain $\left( -\beta ,\beta \right) $ is
found to be 
\begin{equation}
{\cal S}_p=\int_{-\beta }^\beta d\tau \left[ \frac 12m\left( \frac{d\delta _p%
}{d\tau }\right) ^2+U\left( \delta _p\right) \right] =W+2E\beta , 
\eqnum{19a}
\end{equation}
with 
\begin{equation}
W=\frac 83\sqrt{\frac{K_1Vm}\hbar }\left( 1+\frac{27}2\overline{K}_2\right)
\epsilon ^{3/2}\frac 1{\sqrt{1-k^{\prime 2}/2}}\left[ E\left( k\right) -%
\frac{k^{\prime 2}}{2-k^{\prime 2}}K\left( k\right) \right] ,  \eqnum{19b}
\end{equation}
where $k^{\prime 2}=1-k^2$, and $\beta =1/k_BT$ with $k_B$ the Boltzmann
constant. $K\left( k\right) $ and $E\left( k\right) $ are the complete
elliptic integral of the first and second kind, respectively. The general
formula Eq. (4) gives the tunnel splittings of excited levels as $\Delta E=%
\frac{\omega \left( E\right) }\pi \exp \left( -W\right) $, where $W$ is
shown in Eq. (19b), and $\omega \left( E\right) =\frac{2\pi }{t\left(
E\right) }$ is the energy-dependent frequency. For this case, the period $%
t\left( E\right) $ is found to be 
\begin{equation}
t\left( E\right) =\sqrt{2m}\int_{\delta _1}^{\delta _2}\frac{d\delta }{\sqrt{%
E-U\left( \delta \right) }}=2\sqrt{\frac{2\hbar m}{K_1V}}\frac 1{\sqrt{\eta
^2+\alpha }}K\left( k^{\prime }\right) ,  \eqnum{20}
\end{equation}
where $\delta _1=\eta +\sqrt{\eta ^2-\alpha }$, and $\delta _2=\eta +\sqrt{%
\eta ^2+\alpha }$. Now we discuss the low energy limit where $E$ is much
less than the barrier height. In this case, $k^{\prime 4}=\frac{16\hbar E}{%
K_1V\eta ^4}\ll 1$, so we can perform the expansions of $K\left( k\right) $
and $E\left( k\right) $ in Eq. (19b) to include terms like $k^{\prime 4}$
and $k^{\prime 4}\ln \left( \frac 4{k^{\prime }}\right) $, 
\begin{eqnarray*}
E\left( k\right) &=&1+\frac 12\left[ \ln \left( \frac 4{k^{\prime }}\right) -%
\frac 12\right] k^{\prime 2}+\frac 3{16}\left[ \ln \left( \frac 4{k^{\prime }%
}\right) -\frac{13}{12}\right] k^{\prime 4}\cdots , \\
K\left( k\right) &=&\ln \left( \frac 4{k^{\prime }}\right) +\frac 14\left[
\ln \left( \frac 4{k^{\prime }}\right) -1\right] k^{\prime 2}+\frac 9{64}%
\left[ \ln \left( \frac 4{k^{\prime }}\right) -\frac 76\right] k^{\prime
4}\cdots .
\end{eqnarray*}
With the help of small oscillator approximation for energy near the bottom
of the potential well, $E_n=\left( n+\frac 12\right) \Omega _1$, $\Omega _1=%
\sqrt{\frac 1mU^{\prime \prime }\left( \delta =0\right) }=\eta \sqrt{\frac{%
2K_1V}{\hbar m}}$, Eq. (19b) is expanded as 
\begin{equation}
W=W_0-\left( n+\frac 12\right) +\left( n+\frac 12\right) \ln \left[ \frac{1-%
\frac \epsilon 2-\frac{15}2\overline{K}_2}{2^{9/2}S\epsilon ^{3/2}}\left( n+%
\frac 12\right) \right] ,  \eqnum{21a}
\end{equation}
where 
\begin{equation}
W_0=\frac{2^{5/2}}3S\epsilon ^{3/2}\left( 1+\frac \epsilon 2+\frac{15}2%
\overline{K}_2\right) .  \eqnum{21b}
\end{equation}
Then the low-lying energy shift of $n$-th excited states for FM particles
with trigonal crystal symmetry in the presence of an external magnetic field
applied perpendicular to the anisotropy axis $\left( \theta _H=\pi /2\right) 
$ is 
\begin{equation}
\hbar \Delta E_n=\frac 2{n!\sqrt{\pi }}\left( K_1V\right) \epsilon
^{1/2}S^{-1}\left( 1-\frac \epsilon 2+\frac{21}2\overline{K}_2\right) \left( 
\frac{2^{9/2}S\epsilon ^{3/2}}{1-\frac \epsilon 2-\frac{15}2\overline{K}_2}%
\right) ^{n+1/2}\exp \left( -W_0\right) .  \eqnum{22}
\end{equation}
For $n=0$, the energy shift of the ground state is 
\begin{equation}
\hbar \Delta E_0=\frac{2^{13/4}}{\sqrt{\pi }}\left( K_1V\right) \epsilon
^{5/4}S^{-1/2}\left( 1-\frac \epsilon 4+\frac{57}4\overline{K}_2\right) \exp
\left( -W_0\right) .  \eqnum{23}
\end{equation}
Then Eq. (22) can be written as 
\begin{equation}
\hbar \Delta E_n=\frac{q_1^n}{n!}\left( \hbar \Delta E_0\right) , 
\eqnum{24a}
\end{equation}
where 
\begin{equation}
q_1=\frac{2^{9/2}S\epsilon ^{3/2}}{1-\frac \epsilon 2-\frac{15}2\overline{K}%
_2}.  \eqnum{24b}
\end{equation}

Since we have obtained the tunnel splittings at excited levels, it is
reasonable to study the temperature dependence of the tunneling rate. It is
noted that Eqs. (24a) and (24b) are obtained under the condition that the
levels in the two wells are degenerate. In more general cases, the
transition amplitude between two levels separated by the barrier or the
decay rate should be sensitive to this resonance condition for the two
levels. If in the case of the potential with two degenerate levels only one
of the levels is considered as a perturbative metastable state; however, a
fictitious imaginary energy can be calculated by consideration of possible
back and forth tunneling (i.e., by regarding the instanton-antiinstanton
pair as a bounce-like configuration) in the barrier. Therefore there exists
a relation between the level splitting and this imaginary part of metastable
energy level, and has been referred to as the Bogomolny-Fateyev relation
based on equilibrium thermodynamics\cite{15} 
\begin{equation}
\mathop{\rm Im}%
E_n=\pi \left( \Delta E_n\right) ^2/4\omega \left( E_n\right) ,  \eqnum{25}
\end{equation}
where $\omega \left( E_n\right) $ is the frequency of oscillations at energy
level $E_n$. At finite temperature $T$ the decay rate $\Gamma =2%
\mathop{\rm Im}%
E_n$ can be easily found by averaging over the Boltzmann distribution 
\begin{equation}
\Gamma \left( T\right) =\frac 2{Z_0}\sum_n%
\mathop{\rm Im}%
E_n\exp \left( -\hbar E_n\beta \right) ,  \eqnum{26}
\end{equation}
where $Z_0=\sum_n\exp \left( -\hbar E_n\beta \right) $ is the partition
function with the harmonic oscillator approximated eigenvalues $E_n=\left( n+%
\frac 12\right) \Omega _1$. With the help of the Bogomolny-Fateyev relation
Eq. (25), the final result of the tunneling rate at a finite temperature $T$
is found to be 
\begin{equation}
\Gamma \left( T\right) =\frac \pi {2\Omega _1}\left( 1-e^{-\hbar \Omega
_1\beta }\right) \left( \Delta E_0\right) ^2I_0\left( 2q_1e^{-\hbar \Omega
_1\beta /2}\right) ,  \eqnum{27}
\end{equation}
where $\Delta E_0$ and $q_1$ are shown in Eqs. (23) and (24b). $I_0\left(
x\right) =\sum_{n=0}\left( x/2\right) ^{2n}/\left( n!\right) ^2$ is the
modified Bessel function.

Now we discuss briefly the dissipation effect on spin tunneling. For a spin
tunneling problem, it is important to consider the discrete level structure.
It was quantitatively shown that the phenomenon of MQC depends crucially on
the width of the excited levels in the right well.\cite{Garg95} Including
the effects of dissipation, the decay rate, in particular, is given by\cite
{Garg95,Weiss(book),Barbara(review)} 
\begin{equation}
\Gamma _n=\frac 12\left( \Delta E_n\right) ^2\sum_{n^{\prime }}\frac{\Omega
_{nn^{\prime }}}{\left( E_n-E_{n^{\prime }}\right) ^2+\Omega _{nn^{\prime
}}^2},  \eqnum{28}
\end{equation}
where $\Delta E_n$ is the level splitting, $n^{\prime }$ are the levels in
the other well and $\Omega _{nn^{\prime }}$ is the sum of the linewidths of
the $n$th and $n^{\prime }$th levels caused by the coupling of the system to
the environment. For the exact resonance conditions, the temperature
dependence of the decay rate is 
\begin{equation}
\Gamma \left( T\right) =\sum_n\frac{\left( \Delta E_n\right) ^2}{2\Omega _n}%
\exp \left( -\hbar E_n\beta \right) ,  \eqnum{29}
\end{equation}
where the level broadening $\Omega _n$ contains all the details of the
coupling between the magnet and its environment. If the width caused by the
dissipative coupling sufficiently large, the levels overlap, so that the
problem is more or less equivalent to the tunneling into the structureless
continuum. In this case, the results obtained in this paper should be
changed by including the dissipation. It is noted that the purpose of this
paper is to study the spin tunneling at excited levels for single-domain FM
particles in magnetic field at sufficiently low temperatures. Strong
dissipation is hardly the case for single-domain magnetic particles,\cite
{spin-dissipation} and thereby our results are expected to hold. It has been
argued that the decay rate should oscillate on the applied magnetic field
depending on the relative magnitude between the width and the level spacing.%
\cite{11,12,Garg95,Barbara(review),Chudnovsky93} However, it is not clear,
to our knowledge, what should be the effect of finite temperature in the
problem of spin tunneling. The full analysis of spin tunneling onto the
precession levels remains an open problem.

\subsection*{B. $\pi /2<\theta _H<\pi $}

For $\pi /2<\theta _H<\pi $, the critical angle $\theta _c$ is in the range
of $0<\theta _c<\pi /2$, and $\eta \approx \sqrt{2\epsilon /3}$. By applying
the similar method, the problem can be mapped onto a problem of
one-dimensional motion by integrating out $\phi $, and for this case the
effective mass $m$ and the effective potential $U\left( \delta \right) $ in
Eq. (15) are found to be 
\begin{equation}
m=\frac{\hbar S^2\left( 1+\left| \cot \theta _H\right| ^{2/3}\right) }{%
2K_1V\left[ 1-\epsilon +9\overline{K}_2\left( 1+\left| \cot \theta _H\right|
^{2/3}\right) ^{1/2}-\overline{K}_2\frac{3-\left| \cot \theta _H\right|
^{2/3}}{\left( 1+\left| \cot \theta _H\right| ^{2/3}\right) ^{1/2}}+2%
\overline{K}_2\frac{3+\left| \cot \theta _H\right| ^{2/3}}{\left( 1+\left|
\cot \theta _H\right| ^{2/3}\right) ^{3/2}}\right] },  \eqnum{30a}
\end{equation}
and $U\left( \delta \right) =3U_0q^2\left( 1-\frac 23q\right) $, with $%
q=3\delta /2\sqrt{6\epsilon }$, and 
\begin{equation}
U_0=\frac{2^{7/2}}{3^{3/2}}\frac{K_1V}\hbar \epsilon ^{3/2}\frac{\left| \cot
\theta _H\right| ^{1/3}}{1+\left| \cot \theta _H\right| ^{2/3}}\left[ 1-%
\frac{15}2\overline{K}_2\frac 1{\left( 1+\left| \cot \theta _H\right|
^{2/3}\right) ^{1/2}}\right] .  \eqnum{30b}
\end{equation}
The dependence of the effective potential $\overline{E}_1\left( \delta
\right) $ on $\delta \left( =\theta -\theta _0\right) $ for $\theta _H=3\pi
/4$ is plotted in Fig. 2, and $\hbar U\left( \delta \right) =2K_1V\overline{E%
}_1\left( \delta \right) $. The problem now becomes one of MQT, where the
magnetization vector escapes from the metastable state at $\delta =0$
through the barrier by quantum tunneling.

The nonvacuum bounce configuration with an energy $E>0$ is found to be 
\begin{equation}
\delta _p=\frac 23\sqrt{6\epsilon }\left[ a-\left( a-b\right) \text{sn}%
^2\left( \omega _2\tau ,k\right) \right] ,  \eqnum{31}
\end{equation}
where 
\begin{equation}
\omega _2=\frac 12\sqrt{\frac{3U_0}{2m\epsilon }}\sqrt{a-c}.  \eqnum{32}
\end{equation}
$a\left( E\right) >b\left( E\right) >c\left( E\right) $ denote three roots
of the cubic equation 
\begin{equation}
q^3-\frac 32q^2+\frac E{2U_0}=0.  \eqnum{33}
\end{equation}
sn$\left( \omega _2\tau ,k\right) $ is the Jacobian elliptic sine function
of modulus $k=\sqrt{\frac{a-b}{a-c}}$.

The classical action of the nonvacuum bounce configuration Eq. (31) is 
\begin{equation}
{\cal S}_p=\int_{-\beta }^\beta d\tau \left[ \frac 12m\left( \frac{d\delta _p%
}{d\tau }\right) ^2+U\left( \delta _p\right) \right] =W+2E\beta , 
\eqnum{34a}
\end{equation}
with 
\begin{equation}
W=\frac{2^{9/2}}{5\times 3^{3/2}}\sqrt{m\epsilon U_0}\left( a-c\right)
^{5/2}\left[ 2\left( k^4-k^2+1\right) E\left( k\right) -\left( 1-k^2\right)
\left( 2-k^2\right) K\left( k\right) \right] .  \eqnum{34b}
\end{equation}
The period $t\left( E\right) $ of this case is found to be 
\begin{equation}
t\left( E\right) =\sqrt{2m}\int_c^b\frac{d\delta }{\sqrt{E-U\left( \delta
\right) }}=4\sqrt{\frac{2\epsilon m}{3U_0\left( a-c\right) }}K\left(
k^{\prime }\right) ,  \eqnum{35}
\end{equation}
where $k^{\prime 2}=1-k^2$. Then the general formula Eq. (4) gives the
imaginary parts of the metastable energy levels as $%
\mathop{\rm Im}%
E=\frac{\omega \left( E\right) }\pi \exp \left( -W\right) $, where $\omega
\left( E\right) =\frac{2\pi }{t\left( E\right) }$, and $W$ is shown in Eq.
(34b).

Here we discuss the low energy limit of the imaginary part of the metastable
energy levels. For this case, $E_n=\left( n+\frac 12\right) \Omega _2$, $%
\Omega _2=\sqrt{\frac 1mU^{\prime \prime }\left( \delta =0\right) }=\frac 32%
\sqrt{\frac{U_0}{m\epsilon }}$, $a\approx \frac 32\left( 1-\frac{k^{\prime 2}%
}4\right) $, $b\approx \left( \frac 34k^{\prime 2}\right) \left( 1+\frac 34%
k^{\prime 2}\right) $, $c\approx -\frac 34k^{\prime 2}\left( 1+\frac 14%
k^{\prime 2}\right) $, and $k^{\prime 4}=\frac{16E}{27U_0}\ll 1$. After some
calculations, we obtain the imaginary part of the low-lying metastable
excited levels as $\hbar 
\mathop{\rm Im}%
E_n=\frac{q_2^n}{n!}\left( \hbar 
\mathop{\rm Im}%
E_0\right) $, where 
\[
q_2=\frac{2^{25/4}\times 3^{5/4}S\epsilon ^{5/4}\left| \cot \theta _H\right|
^{1/6}}{1-\frac \epsilon 2+\frac 92\overline{K}_2\left( 1+\left| \cot \theta
_H\right| ^{2/3}\right) ^{1/2}+\frac 14\overline{K}_2\frac{2+9\left| \cot
\theta _H\right| ^{2/3}}{\left( 1+\left| \cot \theta _H\right| ^{2/3}\right)
^{1/2}}+\overline{K}_2\frac{3+\left| \cot \theta _H\right| ^{2/3}}{\left(
1+\left| \cot \theta _H\right| ^{2/3}\right) ^{3/2}}}. 
\]
The imaginary part of the metastable ground-state level is 
\begin{eqnarray}
\hbar 
\mathop{\rm Im}%
E_0 &=&\frac{3^{13/9}\times 2^{31/8}}{\sqrt{\pi }}\left( K_1V\right)
\epsilon ^{7/8}S^{-1/2}\frac{\left| \cot \theta _H\right| ^{1/4}}{1+\left|
\cot \theta _H\right| ^{2/3}}\left[ 1-\frac \epsilon 4+\frac 94\overline{K}%
_2\left( 1+\left| \cot \theta _H\right| ^{2/3}\right) \right.  \nonumber \\
&&\left. -\frac 18\overline{K}_2\frac{51-2\left| \cot \theta _H\right| ^{2/3}%
}{\left( 1+\left| \cot \theta _H\right| ^{2/3}\right) ^{1/2}}+\frac 12%
\overline{K}_2\frac{3+\left| \cot \theta _H\right| ^{2/3}}{\left( 1+\left|
\cot \theta _H\right| ^{2/3}\right) ^{3/2}}\right] \exp \left( -W_0\right) .
\eqnum{37a}
\end{eqnarray}
where the WKB exponent is 
\begin{eqnarray}
W_0 &=&\frac{2^{17/4}\times 3^{1/4}}5S\epsilon ^{5/4}\left| \cot \theta
_H\right| ^{1/6}\left[ 1+\frac \epsilon 2-\frac 92\overline{K}_2\left(
1+\left| \cot \theta _H\right| ^{2/3}\right) ^{1/2}\right.  \nonumber \\
&&\left. +\frac 14\overline{K}_2\frac{2+9\left| \cot \theta _H\right| ^{2/3}%
}{\left( 1+\left| \cot \theta _H\right| ^{2/3}\right) ^{1/2}}-\overline{K}_2%
\frac{3+\left| \cot \theta _H\right| ^{2/3}}{\left( 1+\left| \cot \theta
_H\right| ^{2/3}\right) ^{3/2}}\right] .  \eqnum{37b}
\end{eqnarray}
The decay rate at a finite temperature $T$ is found to be 
\begin{equation}
\Gamma \left( T\right) =2%
\mathop{\rm Im}%
E_0\left( 1-e^{-\hbar \Omega _2\beta }\right) \exp \left( q_2e^{-\hbar
\Omega _2\beta }\right) .  \eqnum{38}
\end{equation}

In Fig. 3 we plot the temperature dependence of the tunneling rate for the
typical values of parameters for nanometer-scale single-domain ferromagnets: 
$S=6000$, $\epsilon =1-\overline{H}/\overline{H}_c=0.01,$ $\overline{K}%
_2=0.01$, and $\theta _H=3\pi /4$. From Fig. 3 one can easily see the
crossover from purely quantum tunneling to thermally assisted quantum
tunneling. The temperature $T_0^{\left( 0\right) }$ characterizing the
crossover from quantum to thermal regimes can be estimated as $%
k_BT_0^{\left( 0\right) }=\Delta U/W_0$, where $\Delta U$ is the barrier
height, and $W_0$ is the WKB\ exponent of the ground-state tunneling. It can
be shown that in the cubic potential $\left( q^2-q^3\right) $, the usual
second-order phase transition from the thermal to the quantum regimes occurs
as the temperature is lowered. The second-order phase transition temperature
is given by $k_BT_0^{\left( 2\right) }=\frac{\hbar \omega _b}{2\pi }$, where 
$\omega _b=\sqrt{\frac 1m\left| U^{\prime \prime }\left( x_b\right) \right| }
$ is the frequency of small oscillations near the bottom of the inverted
potential $-U\left( x\right) $, and $x_b$ corresponds to the bottom of the
inverted potential. For the present case, it is easy to obtain that 
\begin{eqnarray*}
k_BT_0^{\left( 2\right) } &=&\frac{2^{1/4}\times 3^{1/4}}\pi \left(
K_1V\right) S^{-1}\epsilon ^{1/4}\frac{\left| \cot \theta _H\right| ^{1/6}}{%
1+\left| \cot \theta _H\right| ^{2/3}}\left[ 1-\frac \epsilon 2+\frac 92%
\overline{K}_2\left( 1+\left| \cot \theta _H\right| ^{2/3}\right)
^{1/2}\right. \\
&&\left. -\frac 14\overline{K}_2\frac{21-2\left| \cot \theta _H\right| ^{2/3}%
}{\left( 1+\left| \cot \theta _H\right| ^{2/3}\right) ^{1/2}}+\overline{K}_2%
\frac{3+\left| \cot \theta _H\right| ^{2/3}}{\left( 1+\left| \cot \theta
_H\right| ^{2/3}\right) ^{3/2}}\right] ,
\end{eqnarray*}
and $k_BT_0^{\left( 0\right) }=\left( 5\pi /18\right) k_BT_0^{\left(
2\right) }\approx 0.87k_BT_0^{\left( 2\right) }$. For a nanometer-scale
single-domain FM particle, the typical values of parameters for the magnetic
anisotropy coefficients are $K_1=10^8$ erg/cm$^3$, and $K_2=10^5$ erg/cm$^3$%
. The radius of the FM particle is about 12 nm and the sublattice spin is $%
10^6$. In Fig. 4, we plot the $\theta _H$ dependence of the crossover
temperature $T_c$ for typical values of parameters for nanometer-scale
ferromagnets at $\epsilon =0.001$ in a wide range of angles $\pi /2<\theta
_H<\pi $. Fig. 4 shows that the maximal value of $T_c$ is about 0.26K at $%
\theta _H=1.76$. The maximal value of $T_c$ as well as $\Gamma $ is expected
to be observed in experiment. If $\epsilon =0.001$, we obtain that $%
T_c\left( 135^{\circ }\right) \backsim $0.23K corresponding to the crossover
from quantum to classical regime. Note that, even for $\epsilon $ as small
as $10^{-3}$, the angle corresponding to an appreciable change of the
orientation of the magnetization vector by quantum tunneling is $\delta _2=%
\sqrt{6\epsilon }$ rad$>4^{\circ }$. It is quite large enough to distinguish
easily between the two states for experimental tests.

\section*{IV. MQC and MQT for tetragonal crystal symmetry}

In this section, we study the FM particles with tetragonal crystal symmetry
in a magnetic field at arbitrarily directed angles in the $ZX$ plane, which
has the magnetocrystalline anisotropy energy 
\begin{equation}
E\left( \theta ,\phi \right) =K_1\sin ^2\theta +K_2\sin ^4\theta
-K_2^{\prime }\sin ^4\theta \cos \left( 4\phi \right) -M_0H_x\sin \theta
\cos \phi -M_0H_z\cos \theta +E_0,  \eqnum{39}
\end{equation}
where $K_1$, $K_2$ and $K_2^{\prime }$ are the magnetic anisotropy
coefficients, and $K_1>0$. In the absence of magnetic field, the easy axes
of this system are $\pm \widehat{z}$ for $K_1>0$. And the field is applied
in the $ZX$ plane as in the previous section. By using the dimensionless
parameters defined in Eq. (8), and choosing $K_2^{\prime }>0$, we find that $%
\phi =0$ is an easy plane for this system, at which Eq. (38) reduces to 
\begin{equation}
\overline{E}\left( \theta ,\phi =0\right) =\frac 12\sin ^2\theta +\left( 
\overline{K}_2-\overline{K}_2^{\prime }\right) \sin ^4\theta -\overline{H}%
\cos \left( \theta -\theta _H\right) +\overline{E}_0,  \eqnum{40}
\end{equation}
where $\overline{K}_2^{\prime }=K_2^{\prime }/2K_1$. Assuming that $\left| 
\overline{K}_2-\overline{K}_2^{\prime }\right| \ll 1$, we obtain the
critical magnetic field and the critical angle as 
\begin{eqnarray}
\overline{H}_c &=&\frac 1{\left[ \left( \sin \theta _H\right) ^{2/3}+\left|
\cos \theta _H\right| ^{2/3}\right] ^{3/2}}\left[ 1+\frac{4\left( \overline{K%
}_2-\overline{K}_2^{\prime }\right) }{1+\left| \cot \theta _H\right| ^{2/3}}%
\right] ,  \nonumber \\
\sin \theta _c &=&\frac 1{\left( 1+\left| \cot \theta _H\right|
^{2/3}\right) ^{1/2}}\left[ 1+\frac 83\left( \overline{K}_2-\overline{K}%
_2^{\prime }\right) \frac{\left| \cot \theta _H\right| ^{2/3}}{1+\left| \cot
\theta _H\right| ^{2/3}}\right] .  \eqnum{41}
\end{eqnarray}
Introducing $\delta \equiv \theta -\theta _0$ ($\left| \delta \right| \ll 1$
in the small $\epsilon $ limit), we derive the energy $\overline{E}\left(
\theta ,\phi \right) $ as 
\begin{equation}
\overline{E}\left( \delta ,\phi \right) =\overline{K}_2^{\prime }\left[
1-\cos \left( 4\phi \right) \right] \sin ^4\left( \theta _0+\delta \right) +%
\overline{H}_x\left( 1-\cos \phi \right) \sin \left( \theta _0+\delta
\right) +\overline{E}_1\left( \delta \right) ,  \eqnum{42}
\end{equation}
where $\overline{E}_1\left( \delta \right) $ is a function of only $\delta $
given by 
\begin{eqnarray}
\overline{E}_1\left( \delta \right) &=&\left[ \frac 12\overline{H}_c\sin
\left( \theta _c-\theta _H\right) +\left( \overline{K}_2-\overline{K}%
_2^{\prime }\right) \sin \left( 4\theta _c\right) \right] \left( \delta
^3-3\delta ^2\eta \right)  \nonumber \\
&&+\left[ \frac 18\overline{H}_c\cos \left( \theta _c-\theta _H\right)
+\left( \overline{K}_2-\overline{K}_2^{\prime }\right) \cos \left( 4\theta
_c\right) \right] \left( \delta ^4-4\delta ^3\eta +6\delta ^2\eta ^2-4\delta
^2\epsilon \right)  \nonumber \\
&&+4\left( \overline{K}_2-\overline{K}_2^{\prime }\right) \epsilon \delta
^2\cos \left( 4\theta _c\right) .  \eqnum{43}
\end{eqnarray}

\subsection*{A. $\theta _H=\pi /2$}

For this case, we obtain that $\eta \approx \sqrt{2\epsilon }\left[
1-4\left( \overline{K}_2-\overline{K}_2^{\prime }\right) \right] $, and $%
\theta _c=\pi /2$. Performing the Gaussian integration over $\phi $, we can
map the spin system onto a problem of particle with effective mass $m$
moving in the one-dimensional potential well $U\left( \delta \right) $. For
this case, 
\[
m=\frac{\hbar S^2}{2VK_1\left[ 1-\epsilon +4\left( \overline{K}_2-\overline{K%
}_2^{\prime }\right) +16\overline{K}_2^{\prime }\right] }, 
\]
and 
\begin{equation}
U\left( \delta \right) =\frac{K_1V}{4\hbar }\left[ 1+12\left( \overline{K}_2-%
\overline{K}_2^{\prime }\right) \right] \delta ^2\left( \delta -2\eta
\right) ^2.  \eqnum{44}
\end{equation}
By applying the method similar to that in Sec. III. A, we obtain the
low-lying tunnel splitting at degenerate excited levels as $\hbar \Delta E_n=%
\frac{q_3^n}{n!}\left( \hbar \Delta E_0\right) $, where $q_3=\frac{%
2^{9/2}S\epsilon ^{3/2}}{1-\frac \epsilon 2+8\left( \overline{K}_2-\overline{%
K}_2^{\prime }\right) +8\overline{K}_2^{\prime }}$. The energy shift of the
ground state is 
\begin{equation}
\hbar \Delta E_0=\frac{2^{13/4}}{\sqrt{\pi }}\left( K_1V\right) \epsilon
^{5/4}S^{-1/2}\left( 1-\frac \epsilon 4+4\overline{K}_2\right) \exp \left(
-W_0\right) .  \eqnum{45a}
\end{equation}
where the WKB exponent is 
\begin{equation}
W_0=\frac{2^{5/2}}3S\epsilon ^{3/2}\left[ 1+\frac \epsilon 2-8\left( 
\overline{K}_2-\overline{K}_2^{\prime }\right) -8\overline{K}_2^{\prime
}\right] .  \eqnum{45b}
\end{equation}
Eqs. (45a) and (45b) agree well with the result obtained with the help of
the vacuum instanton solution.\cite{4} And the final result of the decay
rate at a finite temperature $T$ is $\Gamma \left( T\right) =\left( \Delta
E_0\right) ^2\left[ \pi \left( 1-e^{-\hbar \Omega _3\beta }\right) /2\Omega
_3\right] I_0\left( 2q_3e^{-\hbar \Omega _3\beta /2}\right) $, where $%
I_0\left( x\right) $ is the modified Bessel function.

\subsection*{B. $\pi /2<\theta _H<\pi $}

For $\pi /2<\theta _H<\pi $, $\eta \approx \sqrt{2\epsilon /3}$, the
effective mass $m$ is 
\begin{equation}
m=\frac{\hbar S^2\left( 1+\left| \cot \theta _H\right| ^{2/3}\right) }{%
2K_1V\left[ 1-\epsilon +16\overline{K}_2^{\prime }+\frac 43\left( \overline{K%
}_2-\overline{K}_2^{\prime }\right) \frac{3-2\left| \cot \theta _H\right|
^{2/3}}{1+\left| \cot \theta _H\right| ^{2/3}}\right] },  \eqnum{46a}
\end{equation}
and the effective potential is $U\left( \delta \right) =3U_0q^2\left( 1-%
\frac 23q\right) $, with $q=3\delta /2\sqrt{6\epsilon }$, and 
\begin{equation}
U_0=\frac{2^{7/4}}{3^{3/2}}\frac{K_1V}\hbar \epsilon ^{3/2}\frac{\left| \cot
\theta _H\right| ^{1/3}}{1+\left| \cot \theta _H\right| ^{2/3}}\left[ 1+%
\frac 43\left( \overline{K}_2-\overline{K}_2^{\prime }\right) \frac{%
7-4\left| \cot \theta _H\right| ^{2/3}}{1+\left| \cot \theta _H\right| ^{2/3}%
}\right] .  \eqnum{46b}
\end{equation}
For this case, the imaginary part of the low-lying metastable excited levels
is $\hbar 
\mathop{\rm Im}%
E_n=\frac{q_4^n}{n!}\left( \hbar 
\mathop{\rm Im}%
E_0\right) $, where $q_4=\frac{2^{25/4}\times 3^{5/4}S\epsilon ^{5/4}\left|
\cot \theta _H\right| ^{1/6}}{1-\frac \epsilon 2+8\overline{K}_2^{\prime }+%
\frac 43\left( \overline{K}_2-\overline{K}_2^{\prime }\right) \frac{\left|
\cot \theta _H\right| ^{2/3}-2}{1+\left| \cot \theta _H\right| ^{2/3}}}$.
The imaginary part of the metastable ground-state level is 
\begin{eqnarray}
\hbar 
\mathop{\rm Im}%
E_0 &=&\frac{3^{13/9}\times 2^{31/8}}{\sqrt{\pi }}\left( K_1V\right)
\epsilon ^{7/8}S^{-1/2}\frac{\left| \cot \theta _H\right| ^{1/4}}{1+\left|
\cot \theta _H\right| ^{2/3}}  \nonumber \\
&&\times \left[ 1-\frac \epsilon 4+4\overline{K}_2^{\prime }+\frac 23\left( 
\overline{K}_2-\overline{K}_2^{\prime }\right) \frac{12\left| \cot \theta
_H\right| ^{2/3}-7}{1+\left| \cot \theta _H\right| ^{2/3}}\right] \exp
\left( -W_0\right) .  \eqnum{47a}
\end{eqnarray}
where 
\begin{equation}
W_0=\frac{2^{17/4}\times 3^{1/4}}5S\epsilon ^{5/4}\left| \cot \theta
_H\right| ^{1/6}\left[ 1+\frac \epsilon 2-8\overline{K}_2^{\prime }-\frac 43%
\left( \overline{K}_2-\overline{K}_2^{\prime }\right) \frac{\left| \cot
\theta _H\right| ^{2/3}-2}{1+\left| \cot \theta _H\right| ^{2/3}}\right] . 
\eqnum{47b}
\end{equation}
The final result of the decay rate at a finite temperature $T$ is $\Gamma
\left( T\right) =2%
\mathop{\rm Im}%
E_0\left( 1-e^{-\hbar \Omega _4\beta }\right) \exp \left( q_4e^{-\hbar
\Omega _4\beta }\right) $. And the second-order phase transition temperature
characterizing the crossover from quantum to thermal regimes is found to be 
\begin{eqnarray*}
k_BT_0^{\left( 2\right) } &=&\frac{2^{1/4}\times 3^{1/4}}\pi \left(
K_1V\right) S^{-1}\epsilon ^{1/4}\frac{\left| \cot \theta _H\right| ^{1/6}}{%
1+\left| \cot \theta _H\right| ^{2/3}} \\
&&\times \left[ 1-\frac \epsilon 2+8\overline{K}_2^{\prime }+\frac 43\left( 
\overline{K}_2-\overline{K}_2^{\prime }\right) \frac{5-3\left| \cot \theta
_H\right| ^{2/3}}{1+\left| \cot \theta _H\right| ^{2/3}}\right] .
\end{eqnarray*}

\section*{V. MQC and MQT for hexagonal crystal symmetry}

In this section, we study the hexagonal spin system whose magnetocrystalline
anisotropy energy $E_a\left( \theta ,\phi \right) $ at zero magnetic field
can be written as 
\begin{equation}
E_a\left( \theta ,\phi \right) =K_1\sin ^2\theta +K_2\sin ^4\theta +K_3\sin
^6\theta -K_3^{\prime }\sin ^6\theta \cos \left( 6\phi \right) ,  \eqnum{48}
\end{equation}
where $K_1$, $K_2$, $K_3$, and $K_3^{\prime }$ are the magnetic anisotropic
coefficients. The easy axes are $\pm \widehat{z}$ for $K_1>0$. When we apply
an external magnetic field at an arbitrarily directed angle in the $ZX$
plane, the total energy of this system is given by 
\begin{equation}
E\left( \theta ,\phi \right) =E_a\left( \theta ,\phi \right) -M_0H_x\sin
\theta \cos \phi -M_0H_z\cos \theta +E_0,  \eqnum{49}
\end{equation}
By choosing $K_3^{\prime }>0$, we take $\phi =0$ to be the easy plane, at
which the potential energy can be written in terms of the dimensionless
parameters as 
\begin{equation}
\overline{E}\left( \theta ,\phi =0\right) =\frac 12\sin ^2\theta +\overline{K%
}_2\sin ^4\theta +\left( \overline{K}_3-\overline{K}_3^{\prime }\right) \sin
^6\theta -\overline{H}\cos \left( \theta -\theta _H\right) +\overline{E}_0, 
\eqnum{50}
\end{equation}
where $\overline{K}_3=K_3/2K_1$ and $\overline{K}_3^{\prime }=K_3^{\prime
}/2K_1$.

Under the assumption that $\left| \overline{K}_2\right| $, $\left| \overline{%
K}_3-\overline{K}_3^{\prime }\right| \ll 1$, we obtain the dimensionless
critical field $\overline{H}_c$ and the critical angle $\theta _c$ as 
\begin{eqnarray}
\overline{H}_c &=&\frac 1{\left[ \left( \sin \theta _H\right) ^{2/3}+\left|
\cos \theta _H\right| ^{2/3}\right] ^{3/2}}\left[ 1+\frac{4\overline{K}_2}{%
1+\left| \cot \theta _H\right| ^{2/3}}+\frac{6\left( \overline{K}_3-%
\overline{K}_3^{\prime }\right) }{\left( 1+\left| \cot \theta _H\right|
^{2/3}\right) ^2}\right] ,  \nonumber \\
\sin \theta _c &=&\frac 1{\left( 1+\left| \cot \theta _H\right|
^{2/3}\right) ^{1/2}}\left[ 1+\frac 83\overline{K}_2\frac{\left| \cot \theta
_H\right| ^{2/3}}{1+\left| \cot \theta _H\right| ^{2/3}}+8\left( \overline{K}%
_3-\overline{K}_3^{\prime }\right) \frac{\left| \cot \theta _H\right| ^{2/3}%
}{\left( 1+\left| \cot \theta _H\right| ^{2/3}\right) ^2}\right] . 
\eqnum{51}
\end{eqnarray}
By introducing a small variable $\delta \equiv \theta -\theta _0$ $\left(
\left| \delta \right| \ll 1\text{ in the limit of }\epsilon \ll 1\right) $,
the total energy becomes 
\begin{equation}
\overline{E}\left( \delta ,\phi \right) =\overline{K}_3^{\prime }\left[
1-\cos \left( 6\phi \right) \right] \sin ^6\left( \theta _0+\delta \right) +%
\overline{H}_x\left( 1-\cos \phi \right) \sin \left( \theta _0+\delta
\right) +\overline{E}_1\left( \delta \right) ,  \eqnum{52}
\end{equation}
where $\overline{E}_1\left( \delta \right) $ is a function of only $\delta $
given by 
\begin{eqnarray}
\overline{E}_1\left( \delta \right) &=&\left[ \frac 12\overline{H}_c\sin
\left( \theta _c-\theta _H\right) +\overline{K}_2\sin \left( 4\theta
_c\right) +4\left( \overline{K}_3-\overline{K}_3^{\prime }\right) \left(
5\sin ^3\theta _c\cos ^3\theta _c-3\sin ^5\theta _c\cos \theta _c\right)
\right]  \nonumber \\
&&\times \left( \delta ^3-3\delta ^2\eta \right) +\left[ \frac 18\overline{H}%
_c\cos \left( \theta _c-\theta _H\right) +\overline{K}_2\cos \left( 4\theta
_c\right) +3\left( \overline{K}_3-\overline{K}_3^{\prime }\right) \sin
^2\theta _c\left( \sin ^4\theta _c\right. \right.  \nonumber \\
&&\left. \left. -10\sin ^2\theta _c\cos ^2\theta _c+5\cos ^4\theta _c\right)
\right] \left( \delta ^4-4\delta ^3\eta +6\delta ^2\eta ^2-4\delta
^2\epsilon \right) +\epsilon \delta ^2\left[ 4\overline{K}_2\cos \left(
4\theta _c\right) \right.  \nonumber \\
&&\left. +12\left( \overline{K}_3-\overline{K}_3^{\prime }\right) \sin
^2\theta _c\left( \sin ^4\theta _c-10\sin ^2\theta _c\cos ^2\theta _c+5\cos
^4\theta _c\right) \right] .  \eqnum{53}
\end{eqnarray}

\subsection*{A. $\theta _H=\pi /2$}

For $\theta _H=\pi /2$, i.e., the external magnetic field is applied
perpendicular to the anisotropy axis, we obtain that $\theta _c=\pi /2$ and $%
\eta =\sqrt{2\epsilon }\left[ 1-4\overline{K}_2-12\left( \overline{K}_3-%
\overline{K}_3^{\prime }\right) \right] $. The spin system can be mapped
onto a particle with effective mass $m$ moving in the one-dimensional
potential well $U\left( \delta \right) $, where 
\begin{equation}
m=\frac{\hbar S^2}{2VK_1\left[ 1-\epsilon -4\overline{K}_2-6\left( \overline{%
K}_3-\overline{K}_3^{\prime }\right) -36\overline{K}_3^{\prime }\right] }, 
\eqnum{54a}
\end{equation}
and 
\begin{equation}
U\left( \delta \right) =\frac{K_1V}{4\hbar }\left[ 1+12\overline{K}%
_2+30\left( \overline{K}_3-\overline{K}_3^{\prime }\right) \right] \delta
^2\left( \delta -2\eta \right) ^2.  \eqnum{54b}
\end{equation}
By applying the similar method, we obtain that the energy shift of the $n$%
-th excited level is $\hbar \Delta E_n=\frac{q_5^n}{n!}\left( \hbar \Delta
E_0\right) ,$where 
\[
q_5=\frac{2^{9/2}S\epsilon ^{3/2}}{1-\frac \epsilon 2+8\overline{K}%
_2+24\left( \overline{K}_3-\overline{K}_3^{\prime }\right) +18\overline{K}%
_3^{\prime }}. 
\]
The energy shift of the ground state is 
\begin{equation}
\hbar \Delta E_0=\frac{2^{13/4}}{\sqrt{\pi }}\left( K_1V\right) \epsilon
^{5/4}S^{-1/2}\left[ 1-\frac \epsilon 4-6\left( \overline{K}_3-\overline{K}%
_3^{\prime }\right) +9\overline{K}_3^{\prime }\right] \exp \left(
-W_0\right) ,  \eqnum{55a}
\end{equation}
and the WKB exponent is 
\begin{equation}
W_0=\frac{2^{5/2}}3S\epsilon ^{3/2}\left[ 1+\frac \epsilon 2-8\overline{K}%
_2-24\left( \overline{K}_3-\overline{K}_3^{\prime }\right) -18\overline{K}%
_3^{\prime }\right] .  \eqnum{55b}
\end{equation}
The decay rate at a finite temperature $T$ is 
\[
\Gamma \left( T\right) =\left( \Delta E_0\right) ^2\left[ \pi \left(
1-e^{-\hbar \Omega _5\beta }\right) /2\Omega _5\right] I_0\left(
2q_5e^{-\hbar \Omega _5\beta /2}\right) , 
\]
where 
\[
\Omega _5=2^{3/2}\frac{K_1V}{\hbar S}\epsilon ^{3/2}\left[ 1-\frac \epsilon 2%
+4\overline{K}_2+6\left( \overline{K}_3-\overline{K}_3^{\prime }\right) +18%
\overline{K}_3^{\prime }\right] . 
\]

\subsection*{B. $\pi /2<\theta _H<\pi $}

For this case, the effective mass $m$ and the effective potential $U\left(
\delta \right) $ are 
\[
m=\frac{\hbar S^2\left( 1+\left| \cot \theta _H\right| ^{2/3}\right) }{%
2K_1V\left[ 1-\epsilon +\frac 43\overline{K}_2\frac{3-2\left| \cot \theta
_H\right| ^{2/3}}{1+\left| \cot \theta _H\right| ^{2/3}}+2\left( \overline{K}%
_3-\overline{K}_3^{\prime }\right) \frac{3-4\left| \cot \theta _H\right|
^{2/3}}{\left( 1+\left| \cot \theta _H\right| ^{2/3}\right) ^2}+36\overline{K%
}_3^{\prime }\frac 1{1+\left| \cot \theta _H\right| ^{2/3}}\right] }, 
\]
and 
\begin{eqnarray*}
U\left( \delta \right) &=&\frac{K_1V}\hbar \frac{\left| \cot \theta
_H\right| ^{1/3}}{1+\left| \cot \theta _H\right| ^{2/3}}\left[ 1+\frac 43%
\overline{K}_2\frac{7-4\left| \cot \theta _H\right| ^{2/3}}{1+\left| \cot
\theta _H\right| ^{2/3}}+2\left( \overline{K}_3-\overline{K}_3^{\prime
}\right) \frac{11-16\left| \cot \theta _H\right| ^{2/3}}{\left( 1+\left|
\cot \theta _H\right| ^{2/3}\right) ^2}\right] \\
&&\times \delta ^2\left( \sqrt{6\epsilon }-\delta \right) .
\end{eqnarray*}
The imaginary part of the metastable excited levels is $\hbar 
\mathop{\rm Im}%
E_n=\frac{q_6^n}{n!}\left( \hbar 
\mathop{\rm Im}%
E_0\right) $, and the imaginary part of the ground state is 
\begin{eqnarray}
\hbar 
\mathop{\rm Im}%
E_0 &=&\frac{3^{7/9}\times 2^{31/8}}{\sqrt{\pi }}\left( K_1V\right) \epsilon
^{7/8}S^{-1/2}\frac{\left| \cot \theta _H\right| ^{1/4}}{1+\left| \cot
\theta _H\right| ^{2/3}}\left[ 1-\frac \epsilon 4+\frac 23\overline{K}_2%
\frac{12\left| \cot \theta _H\right| ^{2/3}-7}{1+\left| \cot \theta
_H\right| ^{2/3}}\right.  \nonumber \\
&&\left. 2\left( \overline{K}_3-\overline{K}_3^{\prime }\right) \frac{%
9-13\left| \cot \theta _H\right| ^{2/3}}{\left( 1+\left| \cot \theta
_H\right| ^{2/3}\right) ^2}+9\overline{K}_3^{\prime }\frac 1{1+\left| \cot
\theta _H\right| ^{2/3}}\right] \exp \left( -W_0\right) ,  \eqnum{56a}
\end{eqnarray}
where the WKB exponent is 
\begin{eqnarray}
W_0 &=&\frac{2^{17/4}\times 3^{1/4}}5S\epsilon ^{5/4}\left| \cot \theta
_H\right| ^{1/6}\left[ 1-\frac \epsilon 4+\frac 43\overline{K}_2\frac{%
2-\left| \cot \theta _H\right| ^{2/3}}{1+\left| \cot \theta _H\right| ^{2/3}}%
\right.  \nonumber \\
&&\left. +4\left( \overline{K}_3-\overline{K}_3^{\prime }\right) \frac{%
2-3\left| \cot \theta _H\right| ^{2/3}}{\left( 1+\left| \cot \theta
_H\right| ^{2/3}\right) ^2}-18\overline{K}_3^{\prime }\frac 1{1+\left| \cot
\theta _H\right| ^{2/3}}\right] ,  \eqnum{56b}
\end{eqnarray}
and 
\begin{equation}
q_6=\frac{2^{25/4}\times 3^{5/4}S\epsilon ^{5/4}\left| \cot \theta _H\right|
^{1/6}}{1-\frac \epsilon 2-\frac 43\overline{K}_2\frac{2-\left| \cot \theta
_H\right| ^{2/3}}{1+\left| \cot \theta _H\right| ^{2/3}}-4\left( \overline{K}%
_3-\overline{K}_3^{\prime }\right) \frac{2-3\left| \cot \theta _H\right|
^{2/3}}{\left( 1+\left| \cot \theta _H\right| ^{2/3}\right) ^2}+18\overline{K%
}_3^{\prime }\frac 1{1+\left| \cot \theta _H\right| ^{2/3}}}.  \eqnum{56c}
\end{equation}
The final result of the decay rate at a finite temperature $T$ is $\Gamma
\left( T\right) =2%
\mathop{\rm Im}%
E_0\left( 1-e^{-\hbar \Omega _6\beta }\right) \exp \left( q_6e^{-\hbar
\Omega _6\beta }\right) $, where 
\begin{eqnarray*}
\Omega _6 &=&2^{5/4}\times 3^{1/4}\frac{K_1V}{\hbar S}\epsilon ^{1/4}\frac{%
\left| \cot \theta _H\right| ^{1/6}}{1+\left| \cot \theta _H\right| ^{2/3}}%
\left[ 1-\frac \epsilon 2+\frac 43\overline{K}_2\frac{5-3\left| \cot \theta
_H\right| ^{2/3}}{1+\left| \cot \theta _H\right| ^{2/3}}\right. \\
&&\left. +2\left( \overline{K}_3-\overline{K}_3^{\prime }\right) \frac{%
7-10\left| \cot \theta _H\right| ^{2/3}}{\left( 1+\left| \cot \theta
_H\right| ^{2/3}\right) ^2}+18\overline{K}_3^{\prime }\frac 1{1+\left| \cot
\theta _H\right| ^{2/3}}\right] .
\end{eqnarray*}
The second-order phase transition temperature characterizing the crossover
from quantum to thermal regimes is found to be 
\begin{eqnarray*}
k_BT_0^{\left( 2\right) } &=&\frac{2^{1/4}\times 3^{1/4}}\pi \left(
K_1V\right) S^{-1}\epsilon ^{1/4}\frac{\left| \cot \theta _H\right| ^{1/6}}{%
1+\left| \cot \theta _H\right| ^{2/3}}\left[ 1-\frac \epsilon 2+\frac 43%
\overline{K}_2\frac{5-3\left| \cot \theta _H\right| ^{2/3}}{1+\left| \cot
\theta _H\right| ^{2/3}}\right. \\
&&\left. +2\left( \overline{K}_3-\overline{K}_3^{\prime }\right) \frac{%
7-10\left| \cot \theta _H\right| ^{2/3}}{\left( 1+\left| \cot \theta
_H\right| ^{2/3}\right) ^2}+18\overline{K}_3^{\prime }\frac 1{1+\left| \cot
\theta _H\right| ^{2/3}}\right] .
\end{eqnarray*}

\section*{VI. Conclusions}

In summary, we have theoretically investigated the quantum tunneling of the
magnetization vector between excited levels in single-domain FM
nanoparticles in the presence of an external magnetic field at arbitrary
angle. We consider the FM particles with the general structure of
magnetocrystalline anisotropy. By calculating the nonvacuum instanton in the
spin-coherent-state path-integral representation, we obtain the analytic
formulas for the tunnel splitting between degenerate excited levels and the
imaginary parts of the metastable excited levels in the low barrier limit
for the external magnetic field perpendicular to the easy axis $\left(
\theta _H=\pi /2\right) $, and for the field at an angle between the easy
and hard axes $\left( \pi /2<\theta _H<\pi \right) $. The temperature
dependences of the decay rates are clearly shown for each case. The
low-energy limit of our results agrees well with that of ground-state spin
tunneling. One important conclusion is that the tunneling rate and the
tunnel splitting at excited levels depend on the orientation of the external
magnetic field distinctly. Even a small misalignment of the field with $%
\theta _H=\pi /2$ orientation can completely change the results of the
tunneling rates. Another interesting conclusion concerns the field strength
dependence of the WKB\ exponent in the tunnel splitting or the tunneling
rate. It is found that in a wide range of angles, the $\epsilon \left( =1-%
\overline{H}/\overline{H}_c\right) $ dependence of the WKB exponent is given
by $\epsilon ^{5/4}$ (see Eq. (37b)), not $\epsilon ^{3/2}$ for $\theta
_H=\pi /2$ (see Eq. (21b)). As a result, we conclude that both the
orientation and the strength of the external magnetic field are the
controllable parameters for the experimental test of the phenomena of
quantum tunneling and coherence of the magnetization vector between excited
levels in single-domain FM nanoparticles at sufficiently low temperatures.
If the experiment is to be performed, there are three control parameters for
comparison with theory: the angle of the external magnetic field $\theta _H$%
, the strength of the field in terms of $\epsilon $, and the temperature $T$%
. Furthermore, the $\theta _H$ dependence of the crossover temperature $T_c$
and the angle corresponding to the maximal value of $T_c$ are expected to be
observed in further experiments.

In order to avoid the complications due to distributions of particle size
and shape, some groups have tried to study the temperature and field
dependence of magnetization reversal of individual magnets. Recently,
Wernsdorfer and co-workers have performed the switching field measurements
on individual ferrimagnetic and insulating BaFeCoTiO nanoparticles
containing about $10^5$-$10^6$ spins at very low temperatures (0.1-6K).\cite
{7} They found that above 0.4K, the magnetization reversal of these
particles is unambiguously described by the N\'{e}el-Brown theory of thermal
activated rotation of the particle's moment over a well defined anisotropy
energy barrier. Below 0.4K, strong deviations from this model are evidenced
which are quantitatively in agreement with the predictions of the MQT theory
without dissipation.\cite{3} The BaFeCoTiO nanoparticles have a strong
uniaxial magnetocrystalline anisotropy.\cite{7} However, the theoretical
results presented here may be useful for checking the general theory in a
wide range of systems, with more general magnetic anisotropy. The
experimental procedures on single-domain FM\ nanoparticles of Barium ferrite
with uniaxial symmetry\cite{7} may be applied to the systems with more
general symmetries. Note that the inverse of the WKB exponent $B^{-1}$ is
the magnetic viscosity $S$ at the quantum-tunneling-dominated regime $T\ll
T_c$ studied by magnetic relaxation measurements.\cite{1} Therefore, the
quantum tunneling of the magnetization should be checked at any $\theta _H$
by magnetic relaxation measurements. Over the past years a lot of
experimental and theoretical works were performed on the spin tunneling in
molecular Mn$_{12}$-Ac\cite{20} and Fe$_8$\cite{21,Barbara(review)} clusters
having a collective spin state $S=10$ (in this paper $S=10^3-10^5$). These
measurements on molecular clusters with $S=10$ suggest that quantum
phenomena might be observed at larger system sizes with $S\gg 1$. Further
experiments should focus on the level quantization of collective spin states
of $S=10^2$-$10^4$.

The theoretical calculations performed in this paper can be extended to the
AFM\ particles, where the relevant quantity is the excess spin due to the
small noncompensation of two sublattices. Work along this line is still in
progress. We hope that the theoretical results presented in this paper may
stimulate more experiments whose aim is observing quantum tunneling and
quantum coherence in nanometer-scale ferromagnets.

\section*{Acknowledgments}

R.L. would like to acknowledge Dr. Su-Peng Kou, Dr. Yi Zhou, Professor
Yu-Liang Liu, Professor Zhan Xu and Professor Jiu-Qing Liang for stimulating
discussions. R. L. and J. L. Zhu would like to thank Professor W.
Wernsdorfer and Professor R. Sessoli for providing their paper (Ref. 14). R.
L. is indebted to Dr. Silvia Kleff and Prof. Jan von Delft for many useful
discussions.

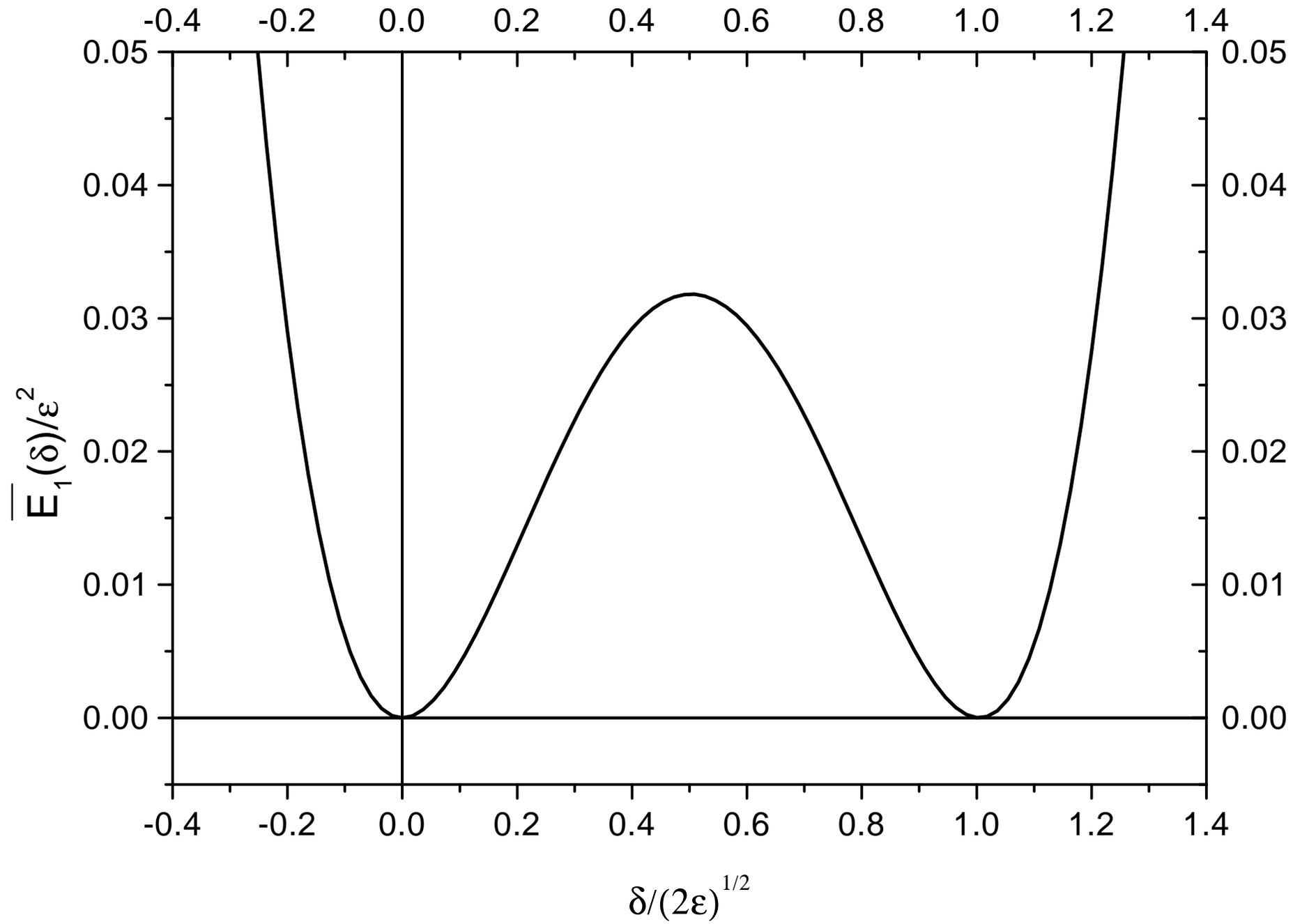
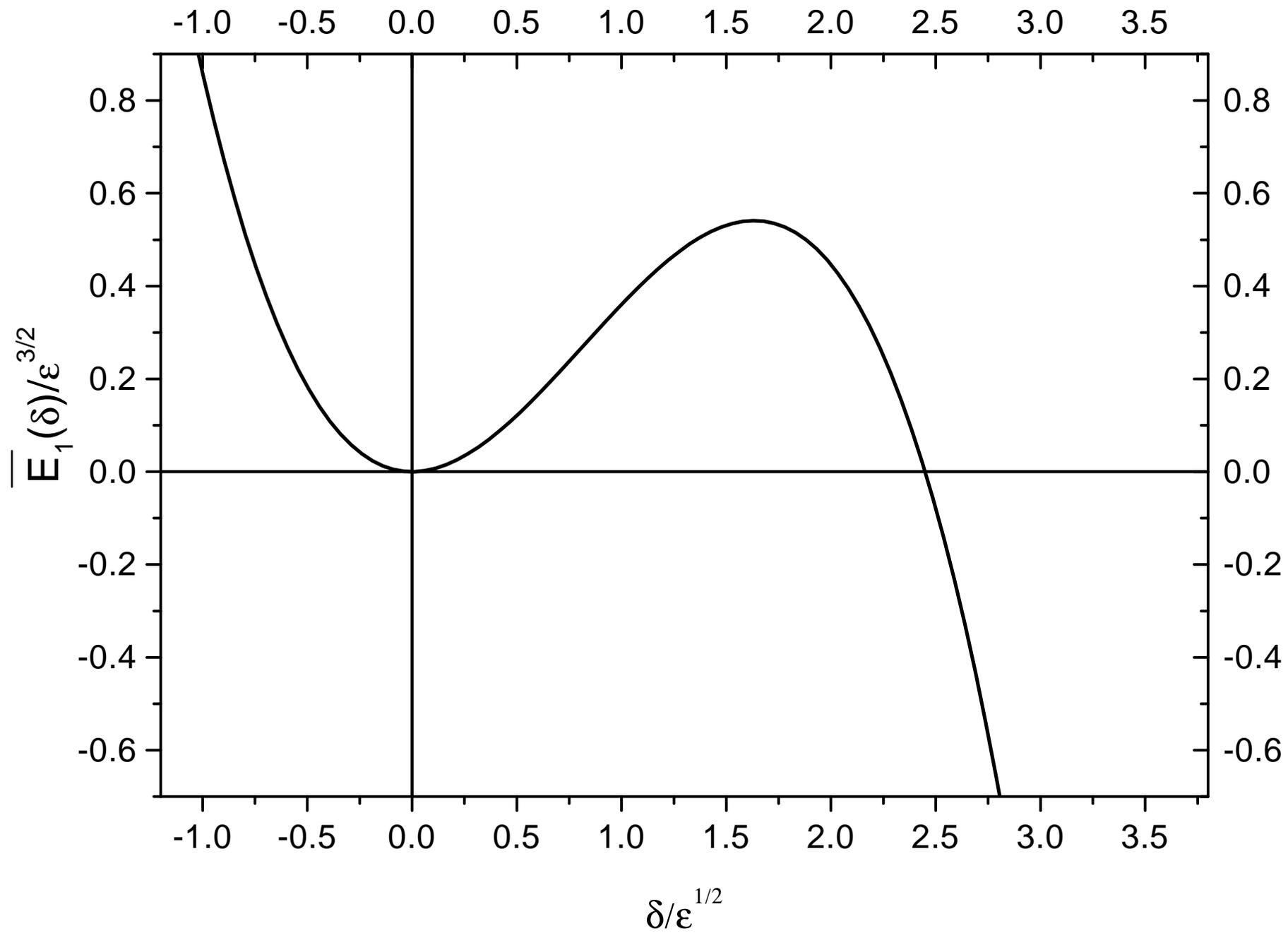
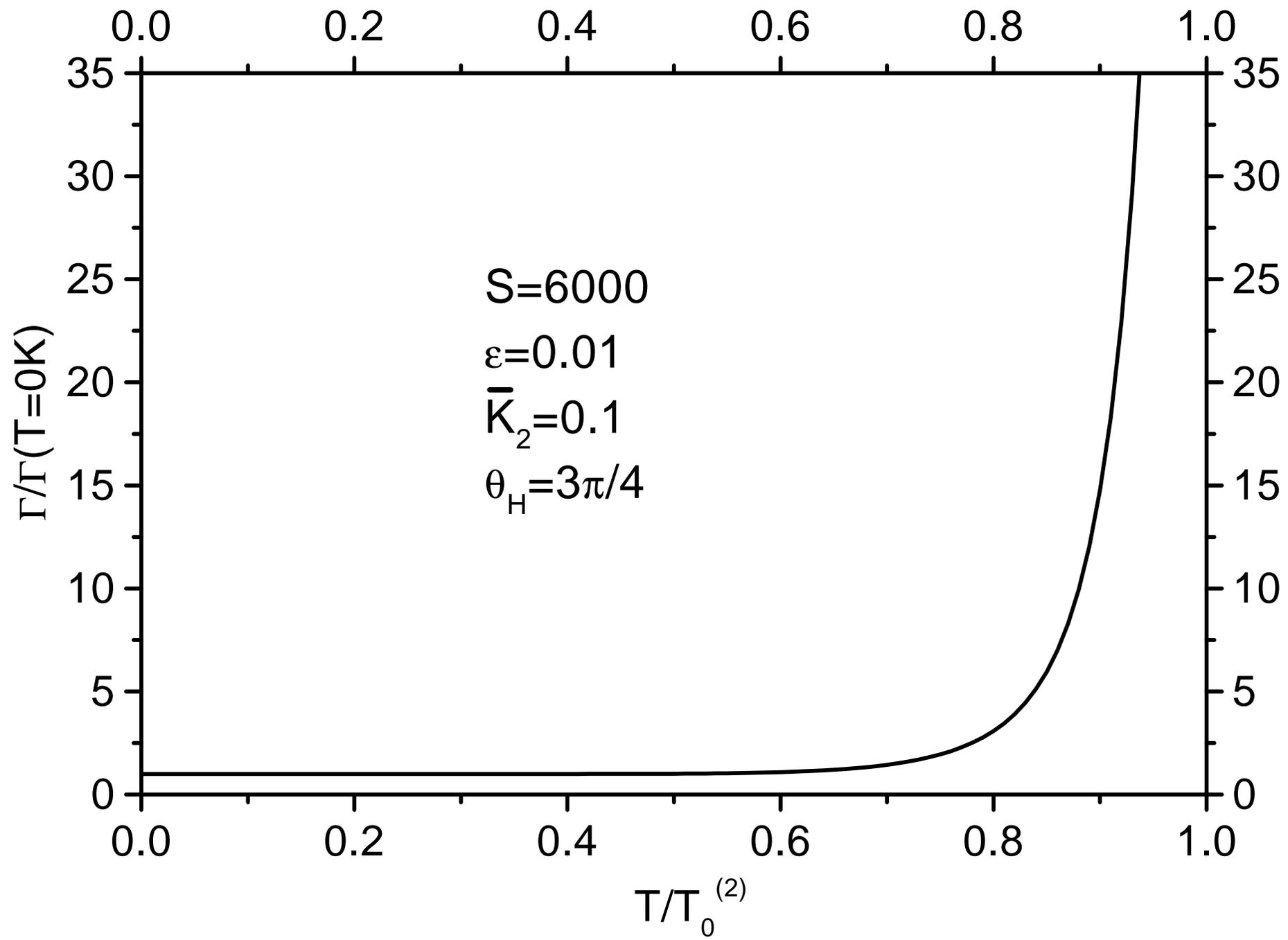
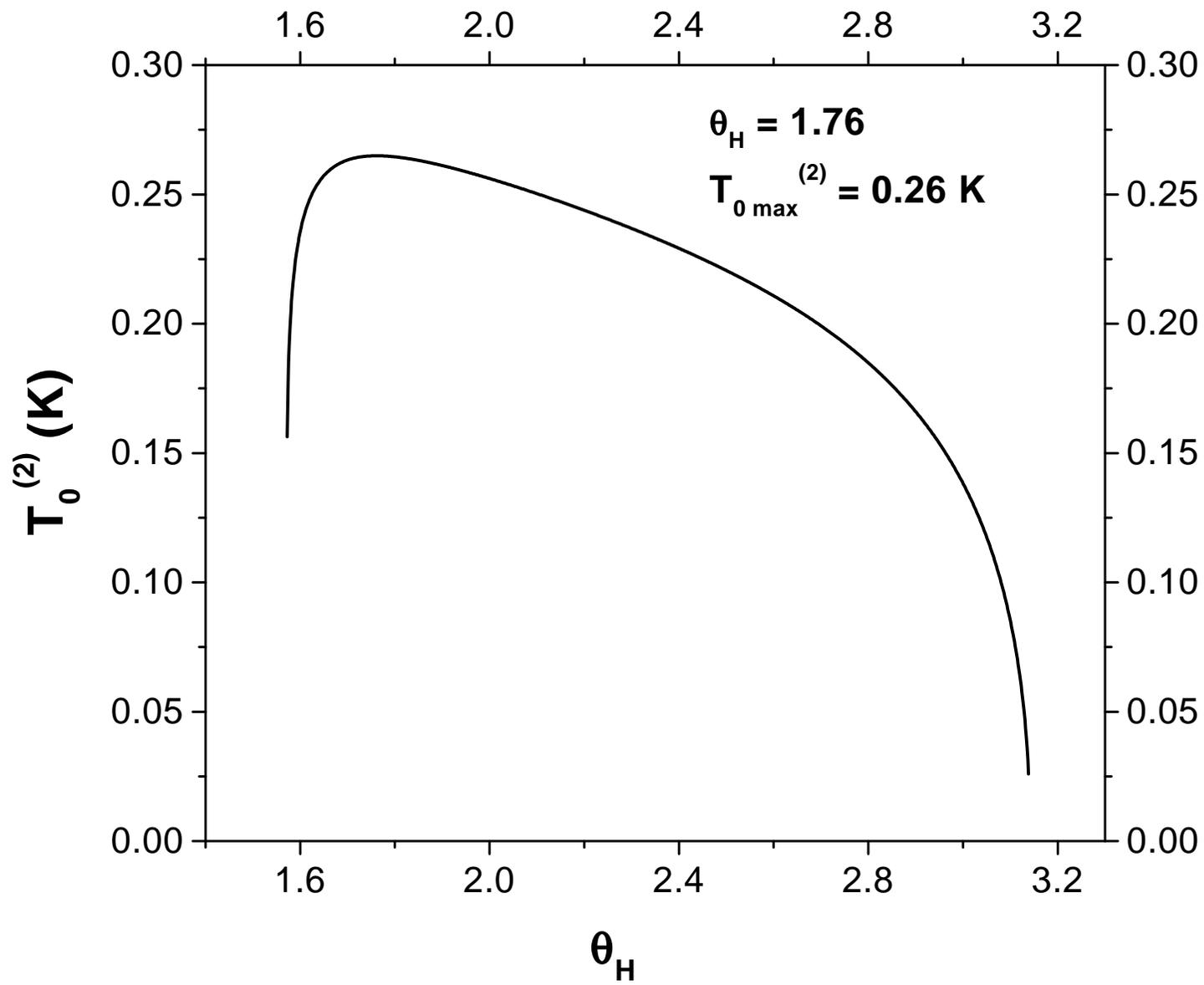
\end{document}